\documentclass[twocolumn,aps,prb,showpacs,floatfix]{revtex4}
\usepackage{graphicx}
\begin{document}
\draft
\title{Pressure dependence of the Boson peak in glasses}
\author{V. L. Gurevich}
\affiliation{Solid State Physics Division, A. F. Ioffe Institute,
194021 Saint Petersburg, Russia}
\author{D. A. Parshin}
\affiliation{Saint Petersburg State Polytechnical
University,
195251 Saint Petersburg, Russia}
\author{H. R. Schober}
\affiliation{Institut f\"ur Festk\"orperforschung, Forschungszentrum
J\"ulich, D-52425 J\"ulich, Germany}
\date{\today}
\begin{abstract}
The inelastic scattering intensities of glasses and amorphous
materials has a maximum at a low frequency, the so called
Boson peak. Under applied hydrostatic pressure, $P$, the Boson peak
frequency, $\omega_{\rm b}$, is shifted upwards. We have shown
previously that the Boson peak is created
as a result of a vibrational instability due to the interaction
of harmonic quasi localized vibrations (QLV).
Applying pressure
one exerts forces on the QLV. These shift the low frequency
part of the excess spectrum to higher frequencies. For low
pressures we find a shift of the Boson peak linear in $P$,
whereas for high pressures the shift is $\propto P^{1/3}$.
Our analytics is supported by simulation. The results are
in agreement with the existing experiments.
\end{abstract}
\maketitle

\section{introduction}
One of the most characteristic properties of glasses is a maximum
in the low frequency part of their inelastic scattering
intensities.~\cite{phil1}
This
maximum, the Boson peak (BP), originates from  a
maximum of the ratio
$g(\omega)/\omega^2$ where $g(\omega)$ is the density of
vibrational states which itself often has no corresponding
maximum.~\cite{ahm1}
The BP shows an excess of low frequency vibrations above the Debye
contribution of the sound waves. The BP
is observed in experiments on
Raman scattering of light and inelastic neutron scattering
within the frequency interval 0.5 --- 2~THz. It is
considered to be one of the universal properties of glasses and is
found also in a number of other disordered systems
(see Ref.~[\onlinecite{GPS:03}] and references therein).

Disorder affects the vibrational states differently than 
it affects the electronic ones. The main point is that the
matrix determining the eigenvalues of the harmonic vibrations,
the squared frequencies, must  satisfy
the requirement of mechanical stability ({\em cf.}
Ref.~\cite{gurarie:02}). In other words, all eigenvalues must
be positive, apart from the six zero values for translation and
rotation of the system as a whole.
Other than in the case of electronic states, there is a fixed
zero level. An arbitrary random matrix has no such
property. This means that, in general, such a matrix
corresponds to an unstable vibrational system.
Such a mechanical instability has often been observed in numerical
simulations of disordered vibrational systems for sufficiently large
degree of disorder ({\it cf.} Refs.~[\onlinecite{schir1,grig1,tar1}]).

However, stability is restored automatically when
the effects of anharmonicity are taken into account.\cite{GPS:03}
A random matrix with the desired stability property is generated in a
natural way by solving the corresponding nonlinear problem.
It is remarkable that the density of states
$g(\omega)$ of such a "stable matrix" possesses the BP feature. The BP is
a reminder of the former mechanical instability in the system.


For the proper interpretation of the BP, the
key problem is the nature of the vibrations that contribute
to $g(\omega)$. Since the term BP is used for any peak in
the low frequency inelastic scattering intensity one has to
distinguish between different cases.
In some materials the BP is
ascribed
to low lying
optical or transverse acoustic modes of parental
crystals~\cite{tezuk1,dove1,sig1}
or to librations of some molecules in plastic
crystals.~\cite{ram,lyn}
If these
excitations have a small frequency spread they will show as
a peak in $g(\omega)$. The role of disorder is merely to broaden
modes which exist already without disorder. In the opposite case the excess
of low frequency modes is caused by disorder itself. A simple example is
realised in a metallic like model glass where the the parent
crystal is fcc.\cite{schober:04}
In the present paper we discuss this latter case.
Between these two extreme cases of well defined low
frequency modes,
broadened by disorder, and no such modes before
disorder,
there is a range of materials having aspects of both.
Despite being derived for the case of a disorder-induced BP,
our results will apply, at least semi-quantitatively, to these
intermediate cases as long as the frequency spread of the
relevant low frequency
modes is  sufficiently large.

Starting point of our investigation are the ubiquitously occurring
quasilocal (or resonant) harmonic vibrations (QLV). These vibrations
can be understood as a low frequency vibration of a small group
of atoms which has a weak bilinear interaction with the continuum
of acoustic
vibrations of the whole system. They share many properties of the
localized vibrations but are different from these exact harmonic
eigenstates. They can be seen as resonances in the low frequency
part
of the local spectra of the set of atoms involved.
In M\"o\ss bauer experiments
one observes anomalously large Debye-Waller factors for the atoms
vibrating with such QLV \cite{petry:82}. In simulations the
QLV are seen in the harmonic eigenstates as ``low frequency localized
modes'' or mixed into the extended modes. This is found for the
textbook case of a heavy mass defect as well as for QLV in glasses
\cite{SR:04}. A ``low frequency localized mode'' of a small system
does not vanish when the system becomes larger -- it just hybridizes
with the other modes of similar frequency. If one looks at the modes
around the QLV-frequency one still finds an additional eigenmode.
A more detailed discussion of QLV with emphasis on glasses can be
found in our previous work on the Boson peak \cite{GPS:03} where one
can
also find supporting references.

In the soft potential model \cite{KKI,IKP} one describes the low
frequency vibrations of the defect system in the harmonic
approximation
utilizing a basis of (extended) sound waves and local oscillators.
The latter are the cores of the QLV or ``bare QLV'' which have been
found to extend over several atoms \cite{BGGS:91,LS:91}. In this
basis there is a bilinear
interaction which is treated as a perturbation.

\section{Effects of finite concentration of quasi localized vibrations}

In a glass one has a finite concentration of QLV's. As discussed in
our previous paper\cite{GPS:03}, this has a profound effect on
their density of states (DOS). The interaction
of the QLV's with the sound waves induces an elastic dipole
interaction
between them. First, the interaction of soft QLV's with surrounding
QLV's of higher frequency may lead in harmonic approximation to
unstable modes. Stability is restored by the anharmonic terms.
This leads to a linear density of states, $g(\omega) \propto \omega$,
for $\omega <  \omega_c$ where $\omega_c$ is determined by the
typical interaction strength.

Secondly, these renormalized low frequency QLV's interact with each
other. In effect this means that the QLV's are subject to random
forces, $f$, later referred to as {\em internal forces}.
Due to the high
susceptibility of low frequency vibrations
their low frequency DOS is changed to $g(\omega) \propto \omega^4$
for $\omega <  \omega_{\rm b}$. This is a general property
of
the low frequency DOS of non-Goldstone bosonic excitations in random media
\cite{gurarie:02} and has been discussed in numerous papers on
low energy excitations in glasses, see
Refs.~[\onlinecite{IKP,BGGS:91,aleiner:94,gurarie:02,fogler:02}].
It is associated with the
so-called {\em sea-gull singularity} in the
distribution of the stiffness constants of the
QLV~\cite{IKP,BGGS:91} in the soft potential model\cite{IKP}.
This sea-gull singularity has also been observed in computer
simulations \cite{heuer:96,SO:96}.

By the crossover between the two limiting
regimes of the DOS the BP is formed and obtains a ``universal''
shape. The frequency of the BP, $\omega_{\rm b}$, is again
determined by the interaction strength and thus by the characteristic
value of the internal forces, $f_0$ (in Ref.~[\onlinecite{GPS:03}]
this quantity was denoted as $\delta f$):
\begin{equation}
\omega_{\rm b} \propto f_0^{1/3}.
\end{equation}
With higher interaction
$\omega_{\rm b}$ is shifted upwards and the intensity of the BP is
reduced \cite{GPS:03}.

One of the most interesting properties of the Boson peak
in glasses is its shift towards higher frequencies under application
of hydrostatic pressure. In the present paper we will show that such
a shift can be visualized as a result of a simple physical mechanism.
If one applies a pressure $P$ onto a specimen of glass the internal
random forces acquire additional random contributions $\Delta f$
proportional to the pressure. As a
result, one obtains a ``blue"
shift of the Boson peak frequency with pressure. For small pressure this
pressure contribution is small compared to the characteristic
value of internal forces $f_0$ and the shift of the BP is linear
in the pressure. With increasing pressure $\Delta f$
can become larger than $f_0$ and the BP shifts as
$\omega_{\rm b}\propto P^{1/3}$.

\section{Random force distribution in a glass under pressure}

In the present Section we will briefly derive, in analogy to
Ref.~[\onlinecite{GPS:03}],
the random force distribution under an
applied hydrostatic pressure. From Hook's law one gets
\begin{equation}
\varepsilon_{ik}=-(P/3K)\delta_{ik}.
\label{1r}
\end{equation}
where $\varepsilon_{ik}$ is the strain tensor and $1/K$ is the
compressibility of the glass.

The interaction of a QLV with the strain is bilinear~\cite{BGGPRS:92}
\begin{equation}
{\cal H}_{\rm int}=\Lambda_{ik}\varepsilon_{ik}
x=-(P/3K)\Lambda_{ii}x
\label{2r}
\end{equation}
where $\Lambda_{ik}$ is the deformation potential tensor and
$x$ is the coordinate of the QLV. For simplicity we will write
in the following $\Lambda$ instead of $\Lambda_{ii}$.
Thus the additional
contribution to the random force due to applied pressure is
proportional to the pressure
\begin{equation}
\Delta f=(P/3K)\Lambda.
\label{3r}
\end{equation}
The deformation potential $\Lambda$ of a QLV is a
random quantity. In particular, it has a random sign, so that
the corresponding distribution function $D(\Lambda)$ is
an even function of $\Lambda$. As a result the distribution of the
random forces $\widetilde{f}$ in the glass remains an
even function of $\widetilde{f}$ when the pressure is applied.

The total random force $\widetilde{f}$ is a sum of two contributions
\begin{equation}
\widetilde{f}=f+\Delta f.
\label{4r}
\end{equation}
Here $f$ is the internal random force in the absence of
pressure. If the distribution of the internal forces $f$
is $Q(f)$
then the distribution of the total random
force $\widetilde{f}$
in a glass under pressure is given by the convolution
\begin{equation}
F_{P}(\widetilde{f})=\int\limits_{-\infty}^{\infty}d\Lambda\,
Q\left(
\widetilde{f} -
\frac{\Lambda}{3K}P\right) D(\Lambda).
\label{5r}
\end{equation}
For $P=0$ it reduces to the unperturbed distribution $Q(f)$ since
the distribution $D(\Lambda)$ is normalized to unity.
In the Appendix we give the results of the convolution for three
cases
which should be typical, namely two types of Lorentzian distributions
and a Gaussian distribution of $f$ and $\Lambda$.

\section{The Boson peak shift under pressure}
\label{shift}

Let us discuss how random forces change the
frequencies of the QLV's. In a purely harmonic
case, the linear forces would not affect the frequencies.
Anharmonicity, however, renormalizes the relevant part of
the spectrum~\cite{IKP,GPS:03}. Although the QLV's are
essentially harmonic vibrations their frequencies under applied
forces can be shifted as in the usual quasi-harmonic approximation.

To start the description of a QLV, consider
an anharmonic oscillator
under the action of a random static
force~$\widetilde{f}$
\begin{equation}
U(x)=Ax^4/4 + M\omega_1^2x^2/2 - \widetilde{f}x.
\label{ao1}
\end{equation}
Here $\omega_1$ is the oscillator frequency in the harmonic
approximation and $A$ is the constant of anharmonicity.
The role of the omitted third order term as well as that of the
distribution of $A$ will be discussed further on.
The force $\widetilde{f}$ shifts the equilibrium
position from
$x=0$ to $x_0\neq 0$, given by
\begin{equation}
\label{xce8}
Ax_0^3 + M\omega_1^2x_0 -
\widetilde{f}
 = 0 ,
\end{equation}
where the oscillator has a new {\em harmonic} frequency
\begin{equation}
\label{kiu6}
\omega^2_{\rm new} = \omega_1^2+3Ax_0^2/M.
\end{equation}

If $\widetilde{g}_1(\omega_1)$ is the distribution function of frequencies
$\omega_1$ and $F_{P}(\widetilde{f})$ is the distribution of random forces
in a glass under pressure, then the pressure dependent
renormalized DOS is given by
\begin{equation}
\label{des2}
g_P(\omega)=\int\limits_0^\infty \widetilde{g}_1(\omega_1)d\omega_1
\int\limits_{-\infty}^{\infty}d
\widetilde{f}
F_{P}(\widetilde{f})\delta\left(\omega -
\omega_{\rm new}\right).
\end{equation}

Integrating the $\delta$-function with regard of Eqs.~(\ref{xce8})
and~(\ref{kiu6}) we get
\begin{equation}
\label{des3}
g_P(\omega)=2M\sqrt{M\over 3A}\omega^3\int\limits_0^\omega
{\widetilde{g}_1(\omega_1)d\omega_1\over\sqrt{\omega^2-\omega_1^2}}
F_P(f_{\omega,\omega_1}).
\end{equation}
Here
\begin{equation}
\label{des4}
f_{\omega,\omega_1}={M\over3}\sqrt{M\over
3A}\,\left(\omega^2+2\omega_1^2\right) \sqrt{\omega^2-\omega_1^2}.
\end{equation}

As shown in Ref.~[\onlinecite{GPS:03}] and shortly discussed in section II
the interaction of high and low frequency QLV lead to a linear DOS
in the relevant frequency range.
In writing Eq.~\ref{ao1} we took QLV already including this effect.
Therefore, we can approximate $\widetilde{g}_1(\omega_1)=C\omega_1$
and write Eq.~(\ref{des3}) in the form


\begin{eqnarray}
\label{des5}
g_P(\omega)=2CM\sqrt{M\over3A}\,
\omega^4\int\limits_0^1{dyy\over\sqrt{1-y^2}}\nonumber\\
\times F_P\left[{M\over3}\sqrt{M\over3A}\,\omega^3\left(1+2y^2\right)
\sqrt{1-y^2}\right].
\end{eqnarray}

Let us introduce a characteristic value $\widetilde{f}_0(P)$ of the
total
random forces acting on the QLV's under pressure. This
is the characteristic scale of variation of the argument of
$F_P(\widetilde{f})$.
Depending on the relative strengths of the internal forces without
pressure and the pressure induced forces, the
following estimates hold
\begin{equation}
\label{des6}
\widetilde{f}_0(P)\approx\left\{{f_0 \left[1 + {\cal O}(P/P_0) \right]
\quad\mbox{for}\quad P\ll P_0
\atop
(P/P_0)f_0\quad\mbox{for}\quad P\gg P_0.}\right.
\end{equation}
Here $P_0$ is the characteristic pressure, which in the simple
case of Lorentzian distributions centered around 0
(see Eqs.~\ref{y5ft}, \ref{eq_D_Lorentz}), is
given by $P_0 =3
Kf_0/\Lambda_0$. Thus for $P\ll P_0$,
$\widetilde{f}_0(P)=f_0$ while for
$P\gg P_0$, $\widetilde{f}_0(P)\propto P$.

According to Ref.~[\onlinecite{GPS:03}], for a Lorentzian distribution
of random forces, the Boson peak frequency is given by
\begin{equation}
\label{1g}
\omega_{\rm b}(P) \approx
{1.9A^{1/6}\widetilde{f}_0^{\,1/3}(P)
\over M^{1/2} } .
\end{equation}

For frequencies $\omega\ll \omega_{\rm b}(P)$ the argument of the
function $F_P$ in Eq.~(\ref{des5}) is much smaller than the
typical value of $\widetilde{f}_0(P)$. One can replace this function by
$F_P(0)\approx 1/\widetilde{f}_0(P)$. According to Eq.~(\ref{des5}),
$g_P(\omega) \propto F_P(0)\omega^4$. Thus at high pressures
$P\gg P_0$,
\begin{equation}
\label{2g}
g_P(\omega)\propto\omega^4/P \hspace{0.5cm}
{\rm for} \quad\omega\ll \omega_{\rm b}(P).
\end{equation}
In the opposite case $\omega\gg \omega_{\rm b}(P)$
the integral over $y$ is dominated by such values of $y$ near
the upper limit that
\begin{equation}
\label{des8}
\sqrt{1-y}\;{{\raise-8pt\hbox{$<$}}\atop\raise6pt\hbox{$\sim$}}\;
\frac{1}{M}\sqrt{\frac{A}{M}}\,\frac{\widetilde{f}_0(P)}{\omega^3}\ll 1.
\end{equation}
After integration, making use of the normalization of $F_P(
\widetilde{f})$,
we regain the equation for the unperturbed pressure-independent
linear density of states
\begin{equation}
\label{yuh5}
g_{P}(\omega)=\widetilde{g}_1(\omega)\propto\omega
\hspace{0.5cm} {\rm for} \quad\omega_{\rm b}(P) \ll \omega < \omega_c
\end{equation}
as it should be.
For higher $\omega$ ($\omega >\omega_c$) the
linear DOS produced by the interaction between the QLV will
be modified and $g_{P}(\omega)$ shows material dependent deviations.

\begin{figure}[htb]
\includegraphics[bb=50 100 460
460,totalheight=6cm,keepaspectratio]{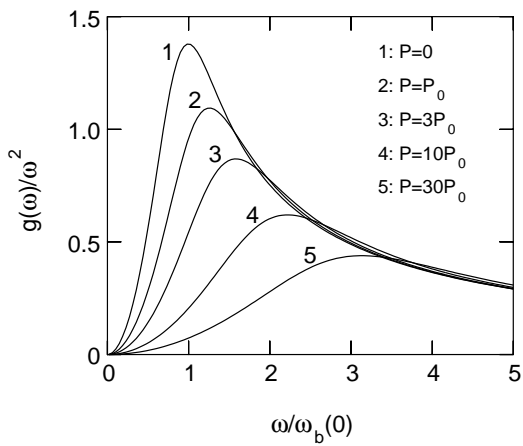}
\caption{Boson peak according to Eq.~(\protect\ref{des5}) for different
applied pressures, excluding
the low frequency Debye contribution. Lorentzian
distributions were assumed for both internal and pressure induced
forces, Eq.~(A3).}
\label{fig_g_om2_P}
\end{figure}

Fig.~\ref{fig_g_om2_P} shows the frequency dependence of the BP for
different applied pressures. Both the distributions of
internal forces $f$ and of the deformation potentials $\Lambda$ were
approximated by Lorentzians centered at 0. After convolution
a broadened Lorentzian is obtained, Eq.~(\ref{7ygt}). One
clearly observes
the pronounced flattening with pressure
of the low frequency part of the BP. Contrary to this, the
high frequency part is not affected. In calculating the curves
in Fig.~\ref{fig_g_om2_P} we assumed $\widetilde{g}_1(\omega) \propto
\omega$. In real materials, this linearity of the DOS will only
hold up to some frequency $\omega_c$. Above that frequency
the DOS will be material dependent and, therefore, the BP will no longer
have a universal form above that frequency. This
can be seen by comparison of the present Fig.~\ref{fig_g_om2_P}
with Fig.~3 of Ref.~[\onlinecite{GPS:03}].

\begin{figure}[htb]
\includegraphics[bb=50 100 460
460,totalheight=6cm,keepaspectratio]{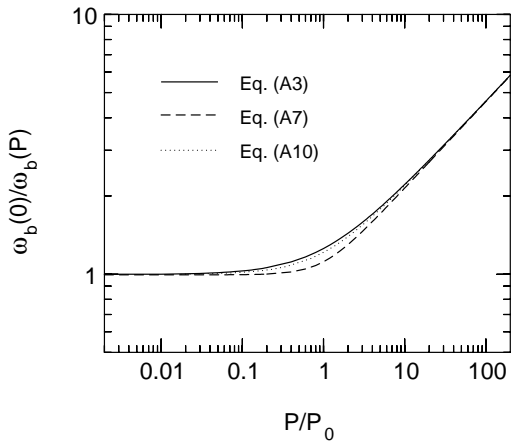}
\caption{Frequency $\omega_{\rm b}$ of the Boson peak maximum as
a function of applied pressure for different force distributions.
Solid line: Lorentzian distributions Eq.~(A3);
dashed line: Gaussian distributions Eq.~(A7);
dotted line: double Lorentzian distribution, Eq.~(A10), with $\lambda =
\protect\sqrt{2}\Lambda_0$.}
\label{fig_omb_P}
\end{figure}
The variation of the BP frequency, $\omega_{\rm b}$, with pressure
is shown in Fig.~\ref{fig_omb_P} for different distributions
of the internal forces and deformation potentials. The
limiting behavior for small and large pressures is independent
of the distributions. The crossover, on the other hand, does
somewhat depend on the type of distribution used. This indicates
a material dependence in this pressure range. The shift of
$\omega_{\rm b}$ with pressure can be described for the Lorentzian
 distributions, Eq.~(A1-A4), by
\begin{equation}
\omega_{\rm b}(P) = \omega_{\rm b}(0)\left(1+\frac{|P|}{P_0} \right)^{1/3}
\label{eq:on6t}
\end{equation}
with
\begin{equation}
P_0=3Kf_0/\Lambda_0 .
\label{eq:u4cr}
\end{equation}
and for the Gaussian distributions, Eq.~(A5-A8), by the
slightly different form
\begin{equation}
\omega_{\rm b}(P) = \omega_{\rm b}(0)\left(1+
\left(\frac{|P|}{P_0}\right)^2 \right)^{1/6}
\label{eq:on6t2g}
\end{equation}
which describes a sharper crossover.

\section{Numerical simulation}
\label{numeric}
\subsection{Pressure dependence}

To test our analytic description we extended the numerical
simulations of Ref.~[\onlinecite{GPS:03}] to include additional external
forces $\Delta f$, Eq.~(\ref{3r}).
We placed $N=2197$
oscillators with frequencies $0<\omega_i<1$ on a simple cubic lattice
with lattice constant $a=1$ and periodic boundary conditions.
The bilinear interaction between two oscillators, $i,j$,
is written as
\begin{equation}
U_{\rm int}^{ij} = g_{ij}\cdot (J/r^3_{ij})\, x_ix_j, 
\label{eq_num_coupl}
\end{equation}
where $r_{ij}$ is the distance between the oscillators and $J$
is the strength of their coupling which results from the
coupling of bare QLV's and sound waves,
$J=\Lambda^2/\rho v^2$.
Here $\rho$ is the density and $v$ the average sound velocity of
the material.
To
simulate random orientations of the oscillators we took for $g_{ij}$
random numbers in the interval $[-0.5,0.5]$.  The
masses, $M_i$, and anharmonicity parameters, $A_i$, were put to 1.
The
DOS for the noninteracting oscillators was taken as $g_0(\omega)
\propto \omega^n$, with
$n=1,2$. Random forces
\begin{equation}
\Delta f^i = g_i \cdot f^0_{\rm ext}
\end{equation} were exerted on the oscillators
where $g_{i}$ were random numbers in the interval $[-0.5,0.5]$.

Generalizing the potential energy
Eq.~(\ref{ao1}) for one oscillator to the system of
$N$ oscillators and adding the
interaction terms described by Eq.~(\ref{eq_num_coupl}), we then
minimized the potential energy of the total system of $N$ coupled
anharmonic oscillators.
In the usual harmonic expansion
around this minimum we calculated the DOS for different values of
$f^0_{\rm ext}$, representing different external pressures.
This was repeated for up to 10000 representations.

\begin{figure}[htb]
\includegraphics[bb=50 100 420
460,totalheight=6cm,keepaspectratio]{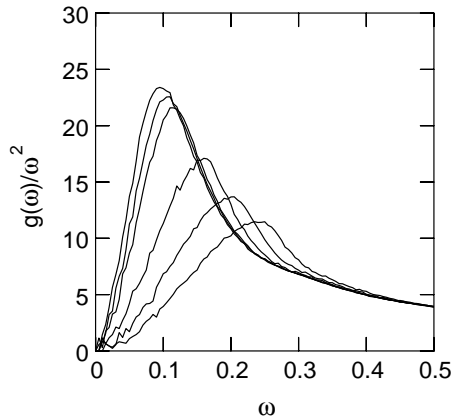}
\caption{Simulated $g(\omega)/\omega^2$ for different external force
strengths ($N=2197$, $g_0(\omega) \propto \omega$, $J=0.07$). Curves
from left to right: $f^0_{\rm ext} = 10^{-4}$, $6\cdot 10^{-4}$,
$10^{-3}$, $3\cdot 10^{-3}$, $6\cdot 10^{-3}$ and $10^{-2}$.}
\label{fig_sim1}
\end{figure}

The frequency dependence of $g(\omega)/\omega^2$ is given in
Fig.~\ref{fig_sim1}, for $g_0(\omega) \propto \omega$ and $J=0.07$,
for different strengths of
the external force, $f^0_{\rm ext}$. The behavior of the analytic
results, Fig.~\ref{fig_g_om2_P}, is reproduced. The slightly
different maximal intensities for the higher forces (pressures)
result from the different choice of distribution for the external
forces (square instead of Lorentzian). The internal forces originate
in the simulation directly from the bilinear interaction.

\begin{figure}[htb]
\includegraphics[bb=50 100 450
460,totalheight=6cm,keepaspectratio]{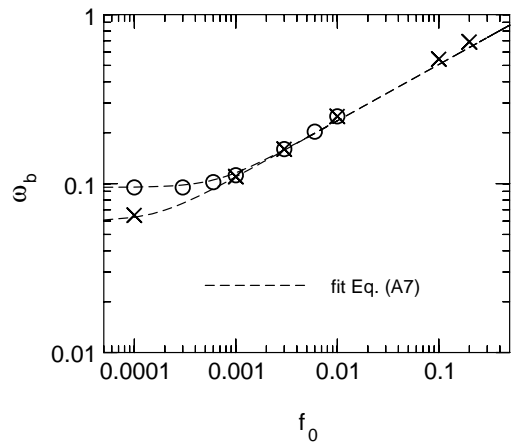}
\caption{Boson peak frequency, $\omega_{\rm b}$, versus external
force strength on a double logarithmic scale: crosses:
$N=2197$, $g_0(\omega) \propto \omega^2$, $J=0.1$; circles:
$N=2197$, $g_0(\omega) \propto \omega$, $J=0.07$. The dashed lines give fits with
Gaussian force distributions (fit parameters $\omega_{\rm b}$ and $P_0$).}
\label{fig_sim2}
\end{figure}

The pressure dependence of BP frequency is shown in a double
logarithmic plot in Fig.~\ref{fig_sim2} for two different
original DOS's: $g_0(\omega) \propto \omega$ and
$g_0(\omega) \propto \omega^2$. The results for both sets
agree with our theoretical predictions. This illustrates that
our results are indeed independent of the choice of the initial
$g_0(\omega)$, as long as it is not too strongly peaked. As
stressed in our previous work, above some frequency $\omega_c$
which is generally well separated above $\omega_{\rm b}$, the
redistribution of frequencies becomes ineffective and the
original $g_0(\omega)$ survives. For instance in a plot
of $g(\omega)/\omega^2$, corresponding to Fig.~\ref{fig_sim1},
for $g_0(\omega) \propto \omega^2$ the curves converge to
a constant given by the normalization, and the maximal intensities
decay more slowly.

\subsection{Oscillator participation numbers}

In our model the formation of the BP is driven by the
interaction between soft oscillators (bare QLV's).
At low frequencies this interaction is weak and, therefore, the
QLV's will be only weakly coupled. The BP frequency is determined
by the typical interaction strength between the
oscillators. To quantify the interaction we introduce an
oscillator participation number
\begin{equation}
n_{\rm osc}(\omega) =  \langle \left( \sum_j |e^j|^4 \right)^{-1}
\rangle_\omega
\label{eq_nosc}
\end{equation}
where $e^j$ denotes the component on oscillator $j$ of an
eigenvector of the coupled oscillator system and
$ \langle ...... \rangle_\omega$ indicates the average over
all eigenmodes of frequency $\omega$. Note that this oscillator
participation number is different from the usual
(atomic or molecular) participation
number of an eigenmode of an atomic system. First a QLV
(an oscillator in the present description) has typically an atomic
participation number of ten or more \cite{LS:91,BGGS:91}. An
oscillator participation number of ten is then equivalent to
an atomic participation number of a hundred or more. Secondly
the participation numbers are further increased by the
interaction between the QLV's and the sound waves \cite{SR:04}.
Here this hybridisation is only included in so far as it
brings about the interaction between QLV's.

\begin{figure}[htb]
\includegraphics[bb=50 100 450
460,totalheight=6cm,keepaspectratio]{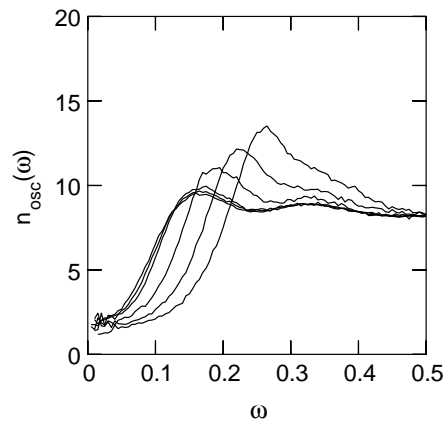}
\caption{Average oscillator participation numbers as function of
frequency for different external force strength. The curves
correspond to the systems of Fig.~\protect\ref{fig_sim1}.}
\label{fig_part}
\end{figure}

For all values of the applied external force, $f^0_{\rm ext}$,
$n_{\rm osc}(\omega)$ shows in Fig.~\ref{fig_part} the same
qualitative behavior as function of $\omega$. For small
frequencies, one has more or less isolated QLV's
($n_{\rm osc}(\omega) \approx 2$). With increasing frequencies,
coupling and hence $n_{\rm osc}(\omega)$ rapidly increases. It
reaches its maximum around $\omega_{\rm b}$ and
drops to a plateau with $n_{\rm osc}(\omega) \approx 8$. This might
at first sight look surprising, since the coupling between
the oscillators, Eq.~(\ref{eq_num_coupl}) was not changed.
On closer inspection of the coupled equations of motion and
the equilibrium condition, one sees, however, that the external
force does in fact change the coupling between the single
oscillators.
The maximal value of $n_{\rm osc}(\omega)$ increases with
$f^0_{\rm ext}$ opposite to $g(\omega_{\rm b})/\omega_{\rm b}^2$.
This is what one would intuitively expect from an increased
coupling. As in the case of the DOS, also $n_{\rm osc}(\omega)$
depends on the original DOS, $g_o(\omega)$, for frequencies 
$\omega >\omega_c > \omega_{\rm b}$.

\subsection{Distribution of anharmonicity parameters}

\begin{figure}[htb]
\includegraphics[bb=50 100 450
460,totalheight=6cm,keepaspectratio]{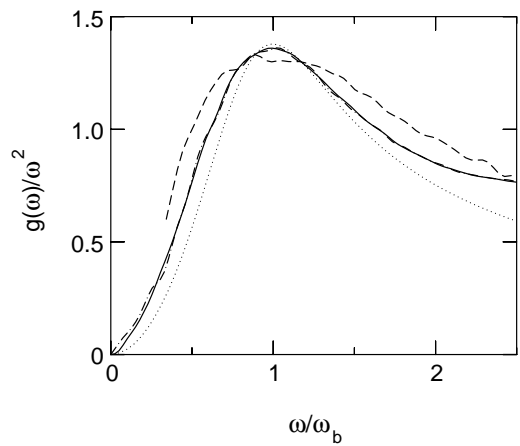}
\caption{Boson peak frequency, $\omega_{\rm b}$, versus external
force strength on a double logarithmic scale: crosses:
$N=2197$, $g_0(\omega) \propto \omega^2$, $J=0.1$; circles:
$N=2197$, $g_0(\omega) \propto \omega$, $J=0.07$. The dashed lines give the
fit with a
Gaussian force distribution.}
\label{fig_anh}
\end{figure}

So far, we have taken the anharmonicity parameter
$A$ in Eq.~\ref{ao1}
as a constant and have neglected possible third order terms.
To check the influence of distributions of these terms, we did
additional simulations where we introduced distributions of
these parameters. The scaled results are summarized in
Fig.~\ref{fig_anh},
where we also show for comparison the theoretical result of
Eq.~\ref{des3} (dotted line). The solid line gives the simulation
results for $g_0(\omega) \propto \omega^2$ and $J=0.2$ with
a fixed value $A=1$. The simulated BP is slightly wider
than the theoretical prediction. The upturn at
$\omega / \omega_{\rm b} \approx 2$ indicates the
upper limit $\omega_c$ where the interaction strength no longer
suffices to destroy the assumed density of the non-interacting
oscillators, $g_0(\omega) \propto \omega^2$. Taking,
for the same parameters, random $A_i$ from the interval
[0.7,1.3], no significant change can be discerned. This is
in agreement with our previous result \cite{GPS:03} that this
anharmonic term provides the mechanism to stabilize the
interacting oscillators but, it's magnitude does not determine
the resulting
spectrum.

The situation is different when we add a third order term to
the energy of the single oscillators,
Eq.~\ref{ao1},
\begin{equation}
U_i(x)=A x^4/4 + B_i x^3/3 + M\omega_i^2x^2/2.
\label{eq_u_anh3}
\end{equation}
The dashed line in Fig.~\ref{fig_anh} shows the resulting BP
for $B_i=b_i \omega_i$ with $b_i$ a random number from
[-1,+1] ($g_0(\omega) \propto \omega$). Compared to the
curves without this term, the BP is considerably broadened even
though it retains its general shape.

\section{Comparison with experiment}
\label{compar}

Unfortunately not too many experimental data are available.
Our theory should, therefore, be considered rather as a prediction
concerning future experiments than as an
interpretation of the existing experimental data.
A general increase of $\omega_{\rm b}$ has been observed in
experiments on a
number of materials, e.g. SiO$_2$ \cite{sugai:96,inamur},
GeO$_2$ \cite{sugai:96}, GeS$_2$ \cite{yamag},
polybutadiene \cite{frick}, polystyrene \cite{geilenkeuser:99}
and teflon \cite{tefl}. Similar shifts have been reported from
computer simulations of SiO$_2$ \cite{jund:00,pilla:03}.
However most of these data are not sufficient for a quantitative
analysis.

\begin{figure}[htb]
\includegraphics[bb=360 580 600
770,angle=-180,totalheight=6cm,keepaspectratio]{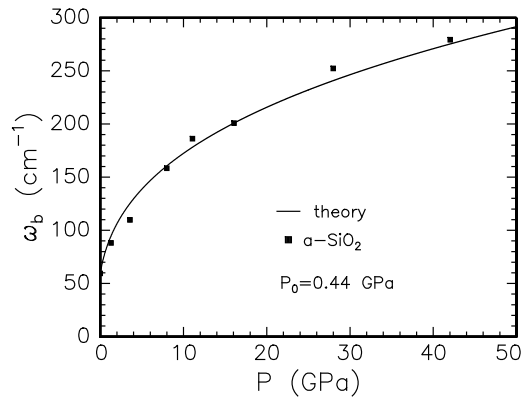}
\caption{The Boson peak in a-SiO$_2$ under
pressure; filled squares are data of Ref.~[\protect\onlinecite{heml}]}
\label{fig:heml1}
\end{figure}

The shift of the BP over a large pressure range has been
measured in  a-SiO$_2$\cite{heml}. As shown in Fig.~\ref{fig:heml1},
the experimental data can be fitted by our theory using
Eqs.~(\ref{eq:on6t}) and~(\ref{eq:u4cr}) assuming a
Lorentzian distribution [see Eq.~(\ref{7ygt})].
The agreement between the theory and experiment remains good even for
high pressures. Regarding very high pressures, our theory is applicable as long as
the short range topology that determines the structure of QLV's
does not change. 

\begin{figure}[htb]
\includegraphics[bb=360 580 600
770,angle=-180,totalheight=6cm,keepaspectratio]{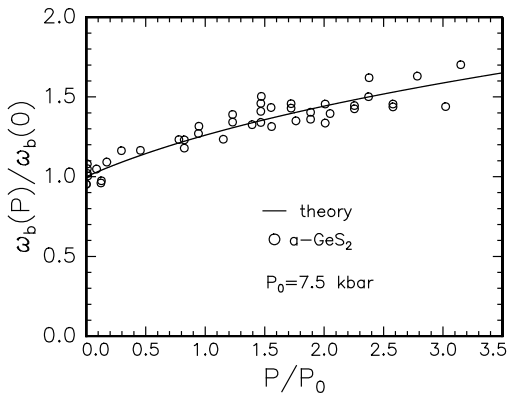}
\caption{The Boson peak in a-GeS$_2$ under
pressure; open circles are data of Ref.~[\protect\onlinecite{yamag}]}
\label{fig:yamag}
\end{figure}

Fig.~\ref{fig:yamag} shows
the shift of the Boson peak in
a-GeS$_2$ measured by Raman scattering.
Although the scatter of the experimental points is rather large,
again the general agreement
with our theory is encouraging.

\begin{figure}[htb]
\includegraphics[bb=360 580 600
770,angle=-180,totalheight=6cm,keepaspectratio]{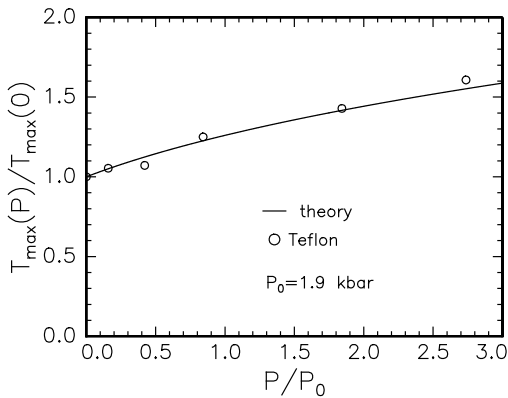}
\caption{The position of the bump $T_{\rm max}(P)$ in the specific
heat
$C(T)/T^3$ in teflon under pressure; open circles are the data of
Ref.~[\protect\onlinecite{tefl}]}
\label{fig:teflbp2}
\end{figure}

Boyer {\it et al.} \cite{tefl} measured the shift with pressure 
of the low temperature maximum  $C(T)/T^3$ in Teflon,
where $C(T)$ is the specific
heat. This maximum is directly related to the BP \cite{buchenau:86}.
Again the observed shift fits well with
our predictions, see Fig.~\ref{fig:teflbp2}.

However, the experiments on the change of the Boson peak
position under pressure are so far insufficient. Therefore, we
believe that
further detailed investigations of this phenomenon are called for.

\section{conclusion}
\label{C}
In our previous paper \cite{GPS:03} we have proposed a mechanism of
the Boson peak formation. The essence of the mechanism is that a
vibrational instability of the spectrum of weakly interacting QLV
is responsible for the origin of the Boson peak in glasses and
some other disordered systems. Anharmonicity stabilizes the
structure but does not determine the shape of the Boson peak.
The vibrations forming the Boson peak are harmonic.

The present paper extends these  ideas.
We show that under the action
of hydrostatic pressure the Boson peak is shifted to higher
frequencies. At comparatively low pressures the shift is linear in
pressure $P$ while for high pressures it is proportional to
$P^{1/3}$.
These conclusions are in good agreement with the existing
experimental data.
Our work explains the shift of the Boson
peek without the need
to postulate additional negative third order anharmonicities
\cite{hizhnyakov:00}.

To obtain a quantitative proof, more extensive investigations of
the pressure dependence
of the Boson peak position in various disordered systems are
needed. Since the proposed mechanism is very general,
it will also be interesting to investigate both theoretically and
experimentally  the behavior of the Boson peak under different
types of
strain other than the hydrostatic one, studied here,
as well as under static
electric fields.
In future work,
we hope to show that the same physical mechanism
is fundamental not only for the formation of the Boson peak but
also for such seemingly different phenomena as creation
of the two level systems that dominate the
properties of glasses at low temperatures.

\acknowledgments
Two of the authors, VLG and DAP, gratefully acknowledge the
financial support
of the German Ministry of Science and Technology and the
hospitality
of the Forschungszentrum J\"ulich where part of the work was done.
VLG also acknowledges the financial support of the Alexander von
Humboldt Foundation and of the Russian Foundation for Basic Research
under grant No 03-02-17638.

\appendix
\section{Convolution}
\label{convol}

To make our investigation more general, we will consider
Lorentzian  and Gaussian distributions for both
the random forces and the deformation potential.
Besides, we will consider also a distribution of the
deformation potential $\Lambda$ that may be called ``double
Lorentzian''. This is formed by two superimposed Lorentzian distributions
with widths $\Lambda_0$ which are centered at $\pm\lambda$, respectively.
For $\lambda>\Lambda_0/\sqrt{3}$
the resulting distribution then 
has two symmetric side maxima and a minimum
at $\Lambda=0$.

For the Lorentzian distributions centered at zero one has
\begin{equation}
Q(f)=\frac{1}{\pi}\frac{f_0}{f^2+f_0^2} = \frac{1}{2\pi}
\int\limits_{-\infty}^\infty d\tau_1e^{if\tau_1-f_0|\tau_1|} .
\label{y5ft}
\end{equation}
and
\begin{equation}
D(\Lambda)=\frac{1}{\pi}\frac{\Lambda_0}{\Lambda^2+\Lambda_0^2}
= \frac{1}{2\pi}
\int\limits_{-\infty}^\infty d\tau_2e^{i\Lambda\tau_2-
\Lambda_0|\tau_2|} .
\label{eq_D_Lorentz}
\end{equation}

As a result of the convolution of these distributions one gets
again a Lorentzian centered at zero but with greater width
\begin{equation}
F_P(f)=\frac{1}{\pi}\frac{\widetilde{f}_0(P)}
{f^2 + \widetilde{f}_0^2(P)}
\label{7ygt}
\end{equation}
where
\begin{equation}
\widetilde{f}_0(P)=f_0+\frac{\Lambda_0 |P|}{3K}.
\label{g3se}
\end{equation}

In the same manner the convolution of two
Gaussian distributions
\begin{equation}
Q(f)=\frac{1}{f_0\sqrt{2\pi}}\,
e^{-f^2/2f_0^2}
= \frac{1}{2\pi} \int\limits_{-\infty}^\infty d\tau_1
e^{if\tau_1-f_0^2\tau_1^2/2} ,
\label{d4rf}
\end{equation}
and
\begin{equation}
D(\Lambda)=\frac{1}{\Lambda_0\sqrt{2\pi}}\,
e^{-\Lambda^2/2\Lambda_0^2}
= \frac{1}{2\pi} \int\limits_{-\infty}^\infty d\tau_2
e^{i\Lambda\tau_2-\Lambda_0^2\tau_2^2/2} ,
\label{dd5f}
\end{equation}
leads to another Gaussian distribution with increased width:
\begin{equation}
F_P(f)=\frac{1}{\widetilde{f}_0(P)
\sqrt{2\pi}}\,
\exp[-f^2/2 \widetilde{f}_0^2(P)]
\label{op8u}
\end{equation}
where
\begin{equation}
\widetilde{f}_0(P)
= \sqrt{f_0^2+\left(\frac{\Lambda_0 P}{3K} \right)^2}.
\label{7ht5}
\end{equation}

Finally let us convolute the Lorentzian for the internal forces,
Eq.~\ref{y5ft}, with a double Lorentzian distribution:
\begin{equation}
D(\Lambda)=\frac{1}{2\pi}\left[\frac{\Lambda_0}{(\Lambda-\lambda)^2+\Lambda_
0^2}+
\frac{\Lambda_0}{(\Lambda+\lambda)^2+\Lambda_0^2}\right].
\label{a1}
\end{equation}
As a result of the convolution one gets
\begin{eqnarray}
F_P(f)=\frac{\widetilde{f}_0(P)}{2\pi}\left[
\frac{1}{(f+\alpha\lambda)^2 + \widetilde{f}_0^2(P)}\right.\nonumber\\
+\left.\frac{1}{(f-\alpha\lambda)^2 + \widetilde{f}_0^2(P)}\right]
\label{2a}
\end{eqnarray}
where $\alpha=P/3K$
and $\widetilde{f}_0(P)$ is given by Eq.~(A4).
This distribution now depends both on the
widths of the two Lorentzians and on the distance between
their centers, $2\lambda$, in $D(\Lambda)$.



\begin{thebibliography}{35}

\bibitem{phil1}
{\it Amorphous Solids. Low Temperature Properties}, edited by W. A.
Phillips, (Springer-Verlag, Berlin:1981).

\bibitem{ahm1}
N. Ahmad, K. W. Hutt, W. A. Phillips, J. Phys. C: Solid State Phys.
{\bf 19}, 3765 (1986).

\bibitem{GPS:03}
V. L. Gurevich, D. A. Parshin and H. R. Schober,
\prb {\bf67}, 094203 (2003).


\bibitem{gurarie:02}V. Gurarie and J. T. Chalker, \prl {\bf 89}, 136801
(2002); \prb {\bf 68}, 134207 (2003).


\bibitem{schir1}
W. Schirmacher, G. Diezemann, C. Ganter, \prl {\bf 81},
136 (1998).

\bibitem{grig1}
T.S. Grigera, V. Martin-Mayor, G. Parisi, P. Verrocchio, J. Phys.:
Cond. Matter {\bf 14}, 2167 (2002).


\bibitem{tar1}
S.N.Taraskin, J.J.Ludlam, G.Natarajan, S.R.Elliott,
Phil.Mag.B {\bf 82}, 197 (2002).

\bibitem{tezuk1}
Y. Tezuka, S. Shin, M. Ishigame, \prl {\bf 66},  2356 (1991).

\bibitem{dove1}
M. T. Dove, M. J. Harris, A. C. Hannon, J. M. Parker, I. P. Swainson,
and M. Gambhir, \prl {\bf 78} 1070 (1997).

\bibitem{sig1}
V. N. Sigaev, E. N. Smelyanskaya, V. G. Plotnichenko, V. V.
Koltashev, A. A. Volkov, and P. Pernice,
J. Non-Cryst. Sol. {\bf 248}, 141 (1999).

\bibitem{ram}
M. A. Ramos, S. Vieira, F. J. Bermejo, J. Dawidowski, H. E. Fischer, H.
Schober, M. A. Gonz\'alez, C. K. Loong, D. L. Price, Phys. Rev.
Lett. {\bf 78}, 82 (1997).

\bibitem{lyn}
R. M. Lynden-Bell, K. H. Michel, Rev. Mod. Phys. {\bf 66}, 721
(1994).

\bibitem{schober:04}
H. R. Schober, J. Phys.: Condens. Matter {\bf 16}, S2659 (2004).

\bibitem{petry:82}
W. Petry, G. Vogl, and W. Mansel, Z. Phys B {\bf 46}, 319 (1982).

\bibitem{SR:04} H. R. Schober and G. Ruocco, Phil. Mag, {\bf 84}, 1361
(2004).

\bibitem{KKI}
V. G. Karpov, M. I. Klinger, and F. N. Ignat'ev, Zh. Eksp. Teor.
Fiz. {\bf 84}, 760 (1983) [Sov. Phys. JETP {\bf 57}, 439 (1983)].

\bibitem{IKP}
M. A. Il'in, V. G. Karpov, D. A. Parshin, Sov. Phys. JETP
{\bf 65}, 165 (1987).

\bibitem{BGGS:91} U. Buchenau, Yu. M. Galperin, V. L. Gurevich, and
H. R. Schober, Phys. Rev. B {\bf 43}, 5039 (1991).

\bibitem{LS:91}
B. B. Laird and H. R. Schober, Phys. Rev. Lett. {\bf 66}, 636
(1991); H. R. Schober and B. B. Laird, Phys. Rev. B {\bf 44},
6746 (1991).

\bibitem{aleiner:94} I. L. Aleiner and I. M. Ruzin, \prl {\bf 72},
1056 (1994).

\bibitem{fogler:02} M. M. Fogler \prl {\bf 88}, 186402 (2002).

\bibitem{heuer:96}
A. Heuer and R. J. Silbey, Phys. Rev. B {\bf 53}, 609 (1996).

\bibitem{SO:96} H. R. Schober and C. Oligschleger, Phys. Rev. B {\bf 53},
11469 (1996).

\bibitem{BGGPRS:92}
U. Buchenau, Yu. M. Galperin, V. L. Gurevich, D. A. Parshin, M. A.
Ramos, H. R. Schober, Phys. Rev. B, {\bf 46}, 2798 (1992).

\bibitem{sugai:96}
S. Sugai and A. Onodera, Phys. Rev. Lett. {\bf 77}, 4210 (1996).

\bibitem{inamur} Y. Inamura,
M. Arai, M. Nakamura, T. Otomo, N. Kitamura, S. M. Bennington,
A. C. Hannon, and U. Buchenau
J.Non-Cryst.Sol. {\bf 293}, 389 (2001).

\bibitem{yamag} M. Yamaguchi, T. Nakayama, T. Yagi, Physica B
{\bf 263-265}, 258 (1999).

\bibitem{frick} B. Frick, C. Alba-Simionesco,
Appl. Phys. A {\bf 74}, S549 (2002).

\bibitem{geilenkeuser:99}
R. Geilenkeuser, Th. Porschberg, M. J\"ackel, and A. Gladun,
Physica B {\bf 263-264}, 276 (1999).

\bibitem{tefl} J. D. Boyer, J. C. Lasjaunias, R. A. Fisher,
and N. E. Phillips, J. Non-Cryst. Sol. {\bf 55}, 413 (1983).

\bibitem{jund:00}
P. Jund and R. Jullien, J. Chem. Phys. {\bf 113}, 2768 (2000).

\bibitem{pilla:03}
O. Pilla, L. Angelani, A. Fontana, J. R. Gon\c{c}alves, and
G. Ruocco, J. Phys.: Condens. Matter {\bf 15}, S995 (2003).

\bibitem{heml} R. J. Hemley, C. Meade, and H. Mao, \prl {\bf 79},
1420 (1997).

\bibitem{buchenau:86}
U. Buchenau, N. N\"ucker, A. J. Dianoux, N. Ahmad, and W. A. Phillips.
Phys. Rev. B {\bf 34}, 5665 (1986).

\bibitem{hizhnyakov:00}
V. Hizhnyakov, A. Laisaar, J. Kikas, An. Kuznetsov, V. Palm,
and A. Suisalu, Phys. Rev. B {\bf 62}, 11296 (2000).





\end{thebibliography}
\end{document}